\documentclass[journal=jacsat,manuscript=article]{achemso}

\usepackage{chemformula} % Formula subscripts using \ch{}
\usepackage[T1]{fontenc} % Use modern font encodings
\usepackage{booktabs}
\usepackage{rotating}
\usepackage{tablefootnote}
\usepackage{array}
\usepackage{multirow}
\usepackage{pdflscape}
\author{Hamidreza Neshasteh}
\affiliation{Matériaux et Phénomènes Quantiques, Université Paris Cité, CNRS-UMR 7162, 75013, Paris, France}
\author{Amideddin Mataji-Kojouri}
\affiliation{Photonics Research Laboratory, School of Electrical and Computer Engineering, University of Tehran, 1439957131, Tehran, Iran}

\author{Clément Le Fur}
\affiliation{Matériaux et Phénomènes Quantiques, Université Paris Cité, CNRS-UMR 7162, 75013, Paris, France}

\author{Ilan Shlesinger}
\affiliation{Matériaux et Phénomènes Quantiques, Université Paris Cité, CNRS-UMR 7162, 75013, Paris, France}

\author{Marco Ravaro}
\affiliation{Matériaux et Phénomènes Quantiques, Université Paris Cité, CNRS-UMR 7162, 75013, Paris, France}

\author{Marc Gély}
\affiliation{Université Grenoble Alpes, CEA, Leti,F-38000, Grenoble,France}
\author{Sébastien Hentz}
\affiliation{Université Grenoble Alpes, CEA, Leti,F-38000, Grenoble,France}
\author{Guillaume Jourdan}
\affiliation{Université Grenoble Alpes, CEA, Leti,F-38000, Grenoble,France}
\author{Ivan Favero}\email{ivan.favero@u-paris.fr}
\affiliation{Matériaux et Phénomènes Quantiques, Université Paris Cité, CNRS-UMR 7162, 75013, Paris, France}

\title[An \textsf{achemso} demo]
  {Multiphysics optomechanical sensing of a liquid on the micron scale}

\abbreviations{IR,NMR,UV}

\begin{document}

%\begin{tocentry}
%
%Some journals require a graphical entry for the Table of Contents.
%This should be laid out ``print ready'' so that the sizing of the
%Inside the \texttt{tocentry} environment, the font used is Helvetica
%8\,pt, as required by \emph{Journal of the American Chemical
%The surrounding frame is 9\,cm by 3.5\,cm, which is the maximum
%permitted for  \emph{Journal of the American Chemical Society}
%graphical table of content entries. The box will not resize if the
%content is too big: instead it will overflow the edge of the box.
%
%This box and the associated title will always be printed on a
%separate page at the end of the document.
%
%\end{tocentry}

\begin{abstract}
  We present an optomechanical device platform for characterization of optical, thermal, and rheological properties of fluids on the micron scale. A suspended silicon microdisk resonator with a vibrating mass of $100 \ fg$ and an effective measurement volume of less than a $\ pL$ is used to monitor %the initial 
  properties of different fluids at rest. 
  By employing analytical models for thermo-optical effects, thermal diffusion and fluid-structure interactions, our platform determines the refractive index, thermal conductivity, viscosity, density and compressibility of the fluid, in a compact measurement setup. A single measurement takes as short as 70 $\mu s$, and the employed power can be less than 100 $\mu W$, guaranteeing measurement at rest and in thermal equilibrium.
\end{abstract}

\section{Introduction and concept}
Liquids are ubiquitous: they form the natural environment for life's mechanisms and are manifesting themselves in diverse industrial processes and cutting-edge printing or bio-technologies. Recent advancements in micro- and nano-fluidic science and applications have spurred an increasing demand for fluid characterization at micro- and nano-metric length scale, aiming at optical, thermal, rheological and other properties of the fluids either "at rest" or "in flow" \cite{puneeth2021microfluidic,del2022review,gu2023simultaneous,salipante2023microfluidic,singh2022comprehensive,toropov2021review,souza2022review,zhou2020viscoelastic}. 

For example, nano-rheometry provides insight into the intricate behavior of biological systems \cite{waigh2016advances,mao2022passive}, shedding light on how cells regulate viscosity for their functions \cite{persson2020cellular,budin2018viscous}. But understanding interdependent biological mechanisms requires actually a concurrent assessment of diverse physical parameters of the biological fluid, not only rheological \cite{gu2023simultaneous}. Integrating different sensing modalities on a single fluidic device platform is hence of prime importance when designing nanofluidic devices for biophysical or biomedical research \cite{trejo2022microfluidics,zhou2020viscoelastic}. For the same reason, it will be crucial as well in other applications such as neuromorphic computing \cite{robin2023long,xiong2023neuromorphic}, chemical synthesis \cite{hou2017interplay,liu2017microfluidics} and in the petroleum, oil and food sectors where micro-rheometry already plays a pivotal role. Of particular importance are measurements on samples with volume down to nanoliters or even picoliters, especially when testing fluids \cite{salipante2023microfluidic} containing proteins \cite{choi2010microfluidic}, DNA, or other scarce or expensive constituents of interest. Optomechanical devices have recently made great strives in this direction, by enabling tracking in real-time a droplet of liquid down to an attoliter volume \cite{Samantha2022}. Scientists across many disciplines have proposed sensors to analyze optical, thermal, rheological, and other properties of fluids. \cite{toropov2021review,salipante2023microfluidic,singh2022comprehensive,del2022review,jimenez2023microfluidic}. 

Most of micro-rheometers reported to date are merely capable of measuring the viscosity and density of fluids \cite{payam2017simultaneous,huang2022piezoelectric,tiwari2023tip,toledo20213d,khan2013online,oliva2019array}. In these devices, the vibrations of a membrane or cantilever at frequencies up to hundreds of kHz are detected through monitoring reflection of a laser beam of a few mW. At variance, micro-rheometers based on ultrasonic transduction operate at a few megahertz and can determine the density and compressibility of a fluid, but extracting viscosity from these measurements is challenging \cite{ledesma2022single}. These probes can operate in passive mode, where the vibrating element is excited by thermal fluctuations \cite{payam2017simultaneous}, or in active mode, for example with piezoelectric \cite{huang2022piezoelectric,tiwari2023tip,toledo20213d,khan2013online} excitation of the vibration. Although active sensors offer superior performance, their practical operation in harsh environments can be constrained by the incompatibility of electrical signals. Optically-operated rheometers are an alternative approach, which employ the photothermal actuation of a membrane or the tracking of microparticles trapped in optical tweezers \cite{oliva2019array,madsen2021ultrafast}. Despite the bulky free-space optical setups and precision alignment that they require, these sensors excel in determining the viscosity \cite{madsen2021ultrafast} along with the density of fluids \cite{oliva2019array}. However, they do not say anything about the compressibility, nor about the optical and thermal properties of the fluid.

Here we report a compact integrated photonic sensor that determines at the same time the optical refractive index, the thermal conductance, as well as the viscosity, density and compressibility (speed of sound) of a fluid. The sensor device is immersed in the fluid and incorporates a small footprint ($16\mu m \times 16 \mu m$) suspended silicon microdisk resonator, which characterizes the fluid within an effective sensing volume of less than a picoliter, within a measurement time as short as 70 $\mu s$. The employed optical power is first kept low at 50 $\mu W$, guaranteeing minimum disturbance to the fluid, and the low power optical response of the disk is used to measure the refractive index of the fluid with a resolution of $4\times 10^{-6} RIU$. By increasing the optical power to 1 mW, we extract the thermal conductivity of the fluid surrounding the sensor with an accuracy better than $3\%$. On top of this optical spectroscopy, the disk serves as well as an optomechanical resonator in the ultra-high frequency range \cite{neshasteh2024optomechanical, gil2015high}, which extracts the fluid viscosity, density and compressibility with high accuracy, by fitting measurements of a given mechanical mode with analytical fluid-structure models \cite{neshasteh2023fluid}. Through simultaneous non-invasive measurement of the optical, thermo-physical and rheological properties of fluids, and measurement time and power orders of magnitude smaller than those of conventional fluid sensors, the demonstrated platform holds promise for diverse applications where multiphysics characterization of liquids on the micron scale is required. 

	Figure 1a is a diagram illustrating the interplay of relevant physical degrees of freedom for the microdisk-fluid environment. The symbols M, O, and T correspond to mechanical, optical, and thermal variables respectively. We show numerical calculations of the velocity associated to the mechanical mode (M), electric field amplitude associated to the optical mode (O), and temperature profile within the disk (T) in Fig. 1b, 1c, and 1d, respectively. Left panels show the cross-section fields and right panels show top views. A bus optical waveguide feeds the optical mode of the disk (O), whose resonant wavelength depends on the fluid refractive index and can be measured in a non-perturbative manner at low power. At higher power, the electromagnetic energy absorbed within the disk resonator (O) results in an increase of the local temperature (T) of both the disk and the fluid in its vicinity. The thermal conductivity of the fluid influences this local temperature variation, which induces variation in the refractive index and spectrally shifts the disk optical mode resonance (O). We can determine the fluid's thermal conductivity by leveraging this thermo-optic effect, through the DC laser spectroscopy of the disk. In parallel, the rheological properties of the fluid impact the mechanical oscillations (M) of the disk, which modulate the optical mode (O) via optomechanical coupling, and are imprinted onto the radio frequency spectrum of the output light. DC and RF components of the detected output light hence provide access to optical, thermal and rheological properties of the liquid.
	
\section{Implementation and results}
A disk is bisected to image with a scanning electron microscope the pedestal holding the structure (In Fig. 1e). A top view of the resonator and its adjacent waveguide is shown in Fig. 1f.  A disk of radius $R=8\ \mu m$ is fabricated on a silicon-on-insulator (SOI) wafer possessing a top silicon layer ($<100>$) of thickness $h=220 \ nm$ above a $1\ \mu m$ thick buried oxide layer. Laser excitation and readout of the microdisk are performed using near-field interaction between the suspended disk and the bus waveguide, which are separated by a 200 nm gap. Grating optical couplers facilitate waveguide input and output coupling. Fig. 1g presents schematics of the measurement setup. The chip hosting the devices is placed within a custom decimeter-sized plastic pool, conveniently fillable with the fluid of interest using a syringe pump. Light from a fiber-coupled tunable laser excites the chip after passing through a fiber polarization controller (FPC). Another optical fiber collects the output light and is subsequently directed to both a low-frequency (DC) and an RF (radio frequency) detection unit \textit{via} an asymmetric beam splitter (BS). Light in the DC path is detected by a slow photodetector, which provides the mean value of optical intensity and allows measuring the refractive index and thermal conductivity of the surrounding fluid. Within the RF path, the light is in contrast amplified by an erbium-doped amplifier (EDFA) before entering a fast photodetector whose electrical output is then amplified by an RF amplifier and analyzed by an electronic spectrum analyzer (ESA). The RF signal is analyzed to extract viscosity, density, and compressibility. We provide here a system of coupled equations that models the behavior of the suspended disk surrounded by a fluid and enable these signal interpretations.\\
Here are the three coupled equations to consider:
%for a quantitative understanding of the evolution of the system behavior and offer valuable insights from the experimental data.
\begin{equation}\label{eq:coupled1}
	\dot{a}+\bigg[\dfrac{\kappa}{2}+j\big[\Delta+ g_{om}x+\dfrac{\omega_c}{n_{eff}}\dfrac{dn_{eff}}{dT} \Delta T \big]\bigg]a=\sqrt{\kappa_{ex}P_{bus}}
\end{equation} 
\begin{equation}\label{eq:coupled2}
	k \nabla^2 T(r) + Q(r) = 0
\end{equation} 
\begin{equation}\label{eq:coupled3}
	(-\omega^2+j\omega\beta+\omega_s^2)\tilde{x}=0
\end{equation}
The first equation rules the evolution of the optical field a in the cavity, with $P_{bus}$ the incident optical power in the adjacent waveguide, $\kappa=\kappa_{abs}+\kappa_{rad}+\kappa_{ex}$ the decay rate of the optical cavity mode, where $\kappa_{abs}$, $\kappa_{rad}$, $\kappa_{ex}$ are the absorption, scattering, and exchange loss rate, respectively. $\Delta=\omega_L-\omega_c$ is the laser to cavity detuning, and $g_{om}=-\frac{d\omega_c}{dx}$ is the parametric optomechanical coupling coefficient. The equation is written for a laser frequency far below the material bandgap, hence neglecting dispersion, such that a mere effective refractive index $n_{eff}$ suffices in the description \cite{Samantha2021}. The second equation rules the heat propagation with the thermal conductivity $k$ and volumetric heat source $Q(r)= \dfrac{P_{abs}}{V}$, where $P_{abs}=\kappa_{abs}\mid a \mid ^2$ is the optical power absorbed within the disk of volume $V$. The third equation shows the equation of mechanical motion in the case of a harmonic oscillation ($x=Re(\tilde{x}e^{j\omega t})$), in absence of Langevin force, and in presence of the fluidic force exerted by a compressible viscous liquid ($\tilde{F}_{fluid}= j\omega m_s\beta \tilde{x}$). $\omega_s^2=\frac{k_s}{m_s}$ is the bare resonant angular frequency with $k_s$ the vibration mode spring constant and $m_s$ the associated mass. $\beta$ can be expressed in complex notation as follows:	
\begin{equation}\label{eq:ch2-m3}
\hspace*{-1.3cm}
	\beta=1+\dfrac{(1+j)\sqrt{2\mu\rho\omega_s}}{h\rho_s}+\dfrac{(1+j)2\sqrt{2\mu\rho\omega_s} + \dfrac{\omega_s\rho h}{2}( \int_{0}^{2KR}J_0(x)dx-j\int_{0}^{2KR}H_0(x)dx)}{a\rho_s(1-\dfrac{J_0(K_sR)J_2(K_sR)}{J_1^2(K_sR)})}+\dfrac{\dfrac{j\omega\rho h}{\pi}ln(\dfrac{32R}{h})}{R\rho_s(1-\dfrac{J_0(K_sR)J_2(K_sR)}{J_1^2(K_sR)})}
\end{equation}	
with $\mu$ the liquid shear viscosity, $\rho$ and $\rho_s$ the density of liquid and solid, $K_s$ the mechanical wave vector in the bare solid resonator, $K$ the acoustic propagation wave vector in liquid $K=\dfrac{\omega_s}{c}$ with c the speed of sound in the liquid, $J_0(x)$ and $H_0(x)$ being Bessel and Struve functions. Let us now describe how these equations can be combined to analyze our experiments and locally obtain the multiphysical properties of the liquid .\\

%Experimental diagram to simultaneously measure optical, thermal, and mechanical properties of liquids within a very small volume is shown in Fig.1 (h).
%Experimental diagram to simultaneously measure optical, thermal, and mechanical properties of liquids within a very small volume is shown in Fig.1 (h). The sample is placed in the cell and immersed in liquid using a syringe pump. The polarization of the tunable laser pump is adjusted with a fiber polarization controller (FPC). The light is coupled into the integrated waveguide via a tilted fiber above the grating and collected and amplified at the output by an
%erbium-doped amplifier (EDFA) followed by an RF amplifier behind the photodetector. The measured signal contains both steady-state and time-dependent information of the optical transmission, therefore, in the optical detection unit the steady-state component of optical transmission is used to measure the refractive index and thermal conductivity of the surrounding fluid. In the RF optomechanical detection unit, the ultrahigh frequency modulation of the optical transmission is measured in conjunction with the mechanical motion of the disk to obtain information about the viscosity, density, and compressibility of the fluid.\\

\begin{figure}[h]%
	\centering
	\includegraphics[width=\textwidth]{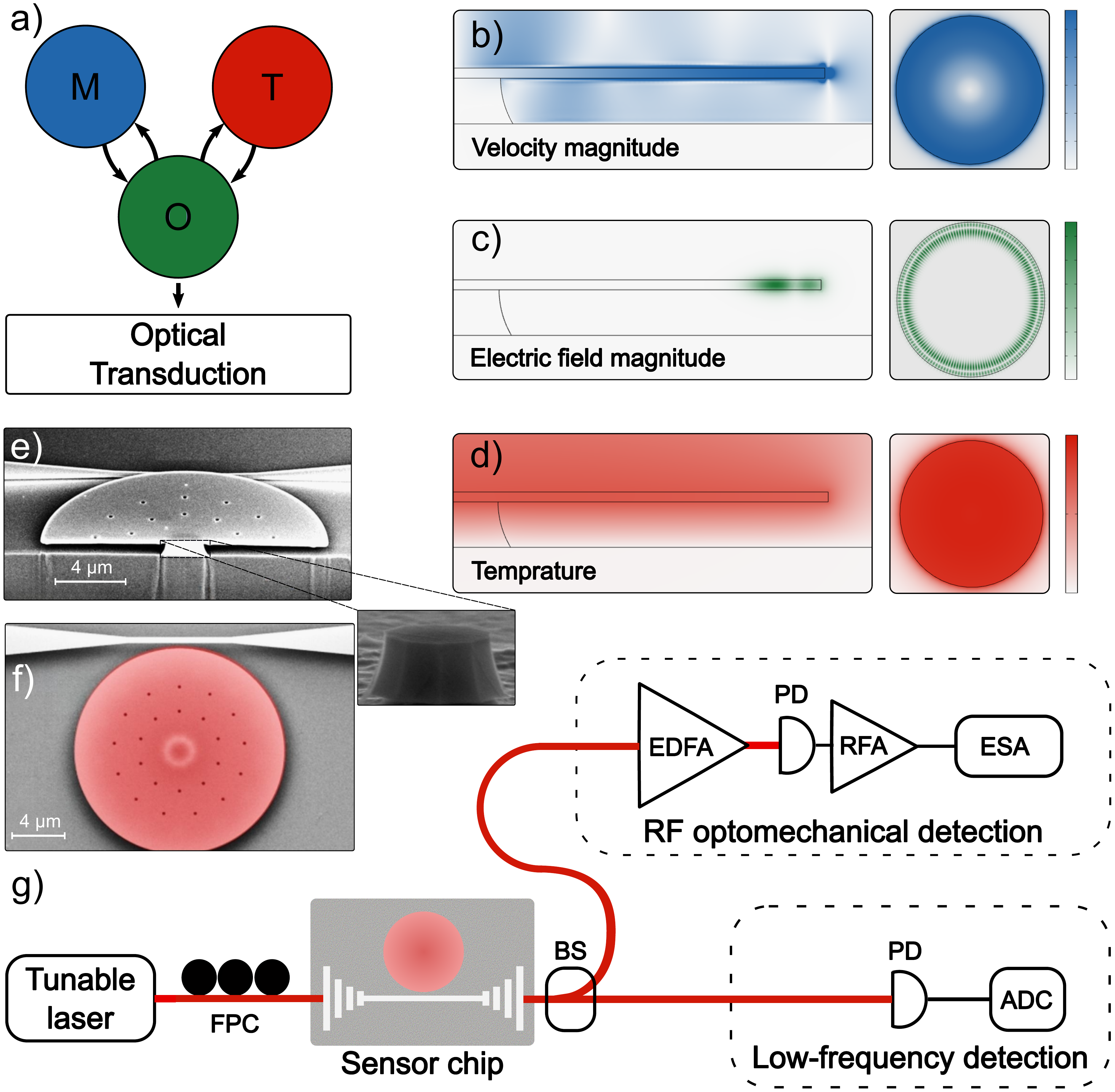}
	\caption{Technological concept of the multi-physics sensor. (a) Schematics of the physics involved. Numerical modal simulations of a suspended silicon disk of radius $8\ \mu m$ and thickness h= 220 nm in water, (b) radial displacement of the fundamental mechanical RBM of the disk, (c) electric field amplitude of an employed optical whispering gallery mode, (d) temperature distribution. Electron micrographs, (e) cross-section, (f) top-view of a suspended silicon disk of radius of $a=8\ \mu m$ and thickness of $h=220 \ nm$, adjacent to a tapered optical bus waveguide. Release holes were drilled through the disk to facilitate underetching of the BOX layer, but play a negligible role in relevant physical interactions. The inset provides a view of the pedestal after etching the top silicon disk. (g) Schematic representation of the measurement setup. PD: Photodetector. BS: Beam splitter. EDFA: Erbium doped fiber amplifier. RFA: Radio frequency amplifier. ESA: Electronic spectrum analyser. ADC: Analog-to-digital converter.}\label{fig1}
\end{figure}
In order to demonstrate that our device and approach can lead to a multiphysical measurement of liquids, several binary water/glycerol mixtures of varying concentration are used as test liquids. %\textcolor{red}{Where are the oils??} 
Table 1 lists the tabulated optical, thermal, and mechanical properties of each binary mixture. Here $n, k, \rho, \mu, c$ stand respectively for the optical
refractive index, thermal conductivity, density, viscosity, and sound velocity of the liquid.\\
\begin{table}[h]
	\centering
	\caption{Optical, thermal and mechanical properties of binary water/glycerol liquid mixtures of varying concentration at $20^\circ c$. Data taken from \cite{hale1973optical,glycerine1963physical,bates1936binary,volk2018density,cheng2008formula,slie1966ultrasonic}.}
	\label{tab:waterGlycerol}
	\begin{tabular}{*8c}
		\toprule
		{$g/w \ (\%)$} & n & {$k \ (\frac{W}{m\cdot K})$}  &{$\rho \ (\frac{kg}{m^3})$}& {$\mu \ (mPa\cdot s)$} & {$c \ (\frac{m}{s})$} \\
		\midrule
		0 & 1.3180 & 0.5894& 997 & 1.05&1510\\
		20 & 1.3474 & 0.5183& 1055&1.87&1728\\
		40 & 1.3759 & 0.4473& 1112&3.79&1844\\
		60 & 1.4050 & 0.3804& 1165&13.96&1909\\	
		80 & 1.4319 & 0.3260& 1214&85.71&1919\\					
		\bottomrule
		\\
		\\
	\end{tabular}
\end{table}\\
Let us first look at the DC optical intensity. By sweeping slowly the wavelength of the tunable laser at low power, the steady-state linear optical response of the device can be obtained. The spectrum calculated  in the steady-state with Eq. \ref{eq:coupled1} depends on the optical properties of the disk and its surrounding environment. The disk's resonance wavelengths ($\lambda_c$) shift when changing the surrounding fluid, depending on the refractive index of the latter ($n$). The optical spectrum of the device exhibits many resonances, associated to distinct families of whispering gallery modes \cite{parrain2015origin} (see Fig. S1 of the Supporting Information). They have various penetration depth into the fluid, which can be used to extend the ability of our sensor to characterize surface adsorbed layers, like in plasmonic biosensing \cite{neshasteh2015hybrid,mataji2020entangled}.
As seen in Fig. 2(a), increasing glycerol concentration increases the refractive index of the liquid mixture and red-shifts a selected optical resonance. The shift in wavelength is linear in a range of more than 0.2 RIU, with a responsivity (S) of $45 nm/RIU$. The minimum detectable refractive index change attains $4\times 10^{-6} RIU$, using $\Delta n_{min}=\Delta \lambda_{min}\times S$, where $\Delta \lambda_{min}= \Delta \lambda_{FWHM}\times F$ \cite{vollmer2012review}, with a measured optical line width of $\Delta \lambda_{FWHM}$=20 pm and F=1/100.\\

By running the same experiment with an increased laser power, we enter the non-linear regime for the optical response. Absorbed optical power within the resonator increases the temperature of the disk, which produces a red-shift of optical resonances, as a consequence of the thermo-optic effect. Solving Eqs. \ref{eq:coupled1} and \ref{eq:coupled2} at steady state requires finding roots of a polynomial, and a solution can be numerically approached in steps to find the optical response \cite{guha2017high}. Fig. 2(b) shows a typical response of the disk at higher power: the response exhibits a non-Lorentzian shape that is consistent with the optical transmission calculated from Eqs. \ref{eq:coupled1} and \ref{eq:coupled2} (see Supporting Information). The exact lineshape depends on the thermal conductivity ($k$) of the liquid, and an increase in thermal conductivity reduces its non-Lorentzian nature.\\

%, while at lower power it remains with a Lorentzian shape, which is perfectly consistent with the expected optical transmission given by the Eq. Eqs. \ref{eq:coupled1}, and .\ref{eq:coupled2}. Since the bistability observed in Fig. 2 (b) depends on the thermal conductivity ($k$) of the liquid, such a probe can be used to measure the thermal properties of the environment.\\
Let us now look at the RF part of the collected signal. Fig. 2(c) shows the power spectral density (PSD) measured at the device output using the RF detection unit, when the disk is immersed in the water-glycerol solutions of distinct concentration. Mechanical resonances of the device appear in such spectrum. The frequency of the first radial breathing mode of the disk in air is 316 MHz, with a modal quality factor of 2200. After immersion in water, the mechanical oscillations are strongly damped by the liquid. With increasing glycerol concentration, the impact of viscous dissipation becomes more pronounced. It is accompanied by a frequency shift of the mechanical vibration towards lower frequency (Fig. 2(c)). The equations (\ref{eq:coupled3}-\ref{eq:ch2-m3}) presented above enable us determining the frequency $\Omega_m=Re(\omega)$ and quality factor $Q_m=\dfrac{Re(\omega)}{2Im(\omega)}$ of the mechanical oscillation, where $\omega$ is the root of Eq. \ref{eq:coupled3}. As seen in Fig. 2(d-e), this model predicts a similar trend for both $Q_m$ and and $\Omega_m$ as a function of glycerol concentration. At high concentration (above $50 \%$), the observed deviation in $\Omega_m$ between model and measurement is attributed to the inability of the highly viscous liquid to penetrate uniformly beneath the disk (gap height $1\ \mu m$). This limitation could be avoided by increasing the height of the pedestal, which requires starting the fabrication from a thicker BOX layer, or by increasing the temperature to let the liquid penetrate this area, before bringing back the temperature at the initial room temperature value. It is worth noting that the sensitivity to the confined region beneath the disk can also offer an opportunity to explore surface tension effects in tight spaces \cite{you2013surface}. At the same time, in contrast to $\Omega_m$, the evolution of $Q_m$ is well captured by our equations, all over the explored range of concentration. Because the raw model of Eqs. (\ref{eq:coupled3}-\ref{eq:ch2-m3}) for mechanical oscillations does not fully reproduce the measured behavior at the highest glycerol concentration, we developed an algorithm to calibrate our model with data measured in air and in water (see the Supporting Information).
%the model deviates from the measurement results, which can be due to the fact that the highly viscous liquid can no longer form a homogeneous layer in the micron volume near the solid interface.

\begin{figure}[h]%
	\centering
	\includegraphics[width=0.9\textwidth]{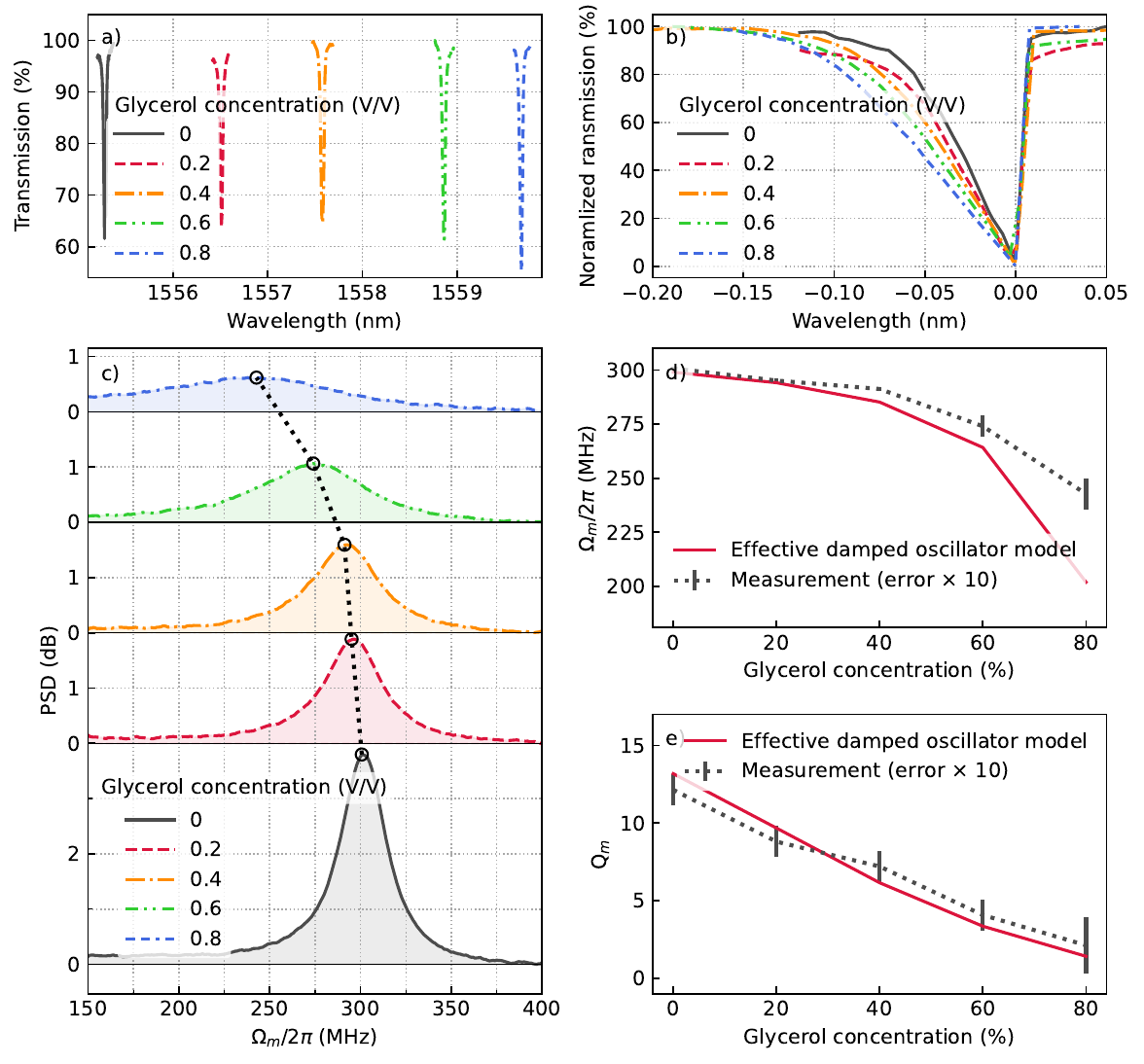}
	\caption{Tracking the evolution of the DC and RF components of the optical signal collected at the output of the bus waveguide. The waveguide is positioned 200 nm away from a silicon disk of radius $8\ \mu m$ and thickness h = 220 nm that is immersed in a binary mixture of water/glycerol of varying concentration. The temperature is kept at 20 °C. (a) DC optical waveguide transmission spectra at low injected optical power. (b) DC optical transmission spectra at higher injected optical power into the waveguide ($P_{bus}=200\ \mu W$). (c) Measured power spectral density (PSD) around the 1st radial breathing mode (RBM) ($P_{bus}= 200\ \mu W$). Measurements and raw model calculations for the RBM center frequency (d) and quality factor (e).}\label{fig2}
\end{figure}

\section{Discussion}

\begin{figure}[h]%
	\centering
	\includegraphics[width=\textwidth]{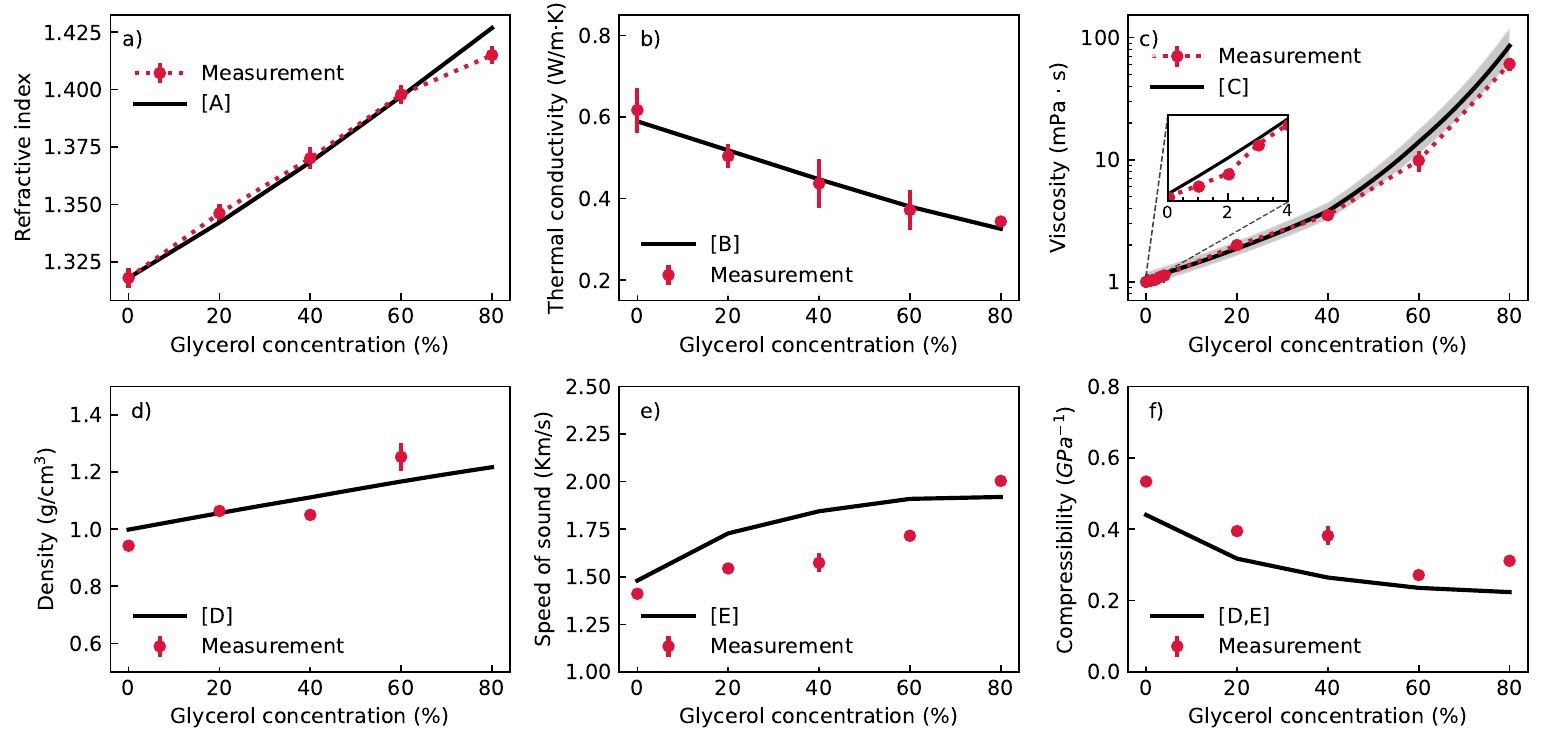}
	\caption{Optical, thermal, and rheological fluid properties deduced from the fit of measurements (red circles with error bar) and comparison with values found in the literature for binary mixtures of water/glycerol of varying concentration. (a) Refractive index, (b) thermal conductivity, (c) viscosity, (d) density, (e) speed of sound, (f) compressibility. [A]\cite{hale1973optical,glycerine1963physical}. [B]\cite{bates1936binary}. [C]\cite{cheng2008formula}. [D]\cite{volk2018density}. [E]\cite{slie1966ultrasonic}.
	}\label{fig3}
\end{figure}

With this calibration of the model now implemented, the optical refractive index, thermal conductivity, viscosity, density and compressibility of the fluid are derived from the results of the low-frequency and RF optomechanical detection units. Extracted physical parameters for different water-glycerol solutions are shown in Fig. \ref{fig3}. As illustrated in Fig. \ref{fig3}a and b, the measured refractive index and thermal conductivity show a high level of agreement with tabulated values, with accuracy better than $0.01 \% $ and $3 \%$, respectively. At the same time, the resolution in the measurement of these two quantities is of $4\times 10^{-6}$ and $6\times 10^{-4}$. In the viscosity data shown in Fig. \ref{fig3}c (inset), the accuracy is better than $2 \%$ for low ($<20\%$) concentration of glycerol in water. However, at higher concentration it is only of $30 \%$. In Fig. \ref{fig3}d-f, we present the extracted density, sound velocity and compressibility ($1/\rho c^2$) of the liquid mixtures. The accuracy is better than $10 \%$ for these parameters, and the resolution of the measurement is also of $10 \%$.
We observe that our optomechanical approach provides lower accuracy and resolution in the measurement of the mechanical properties of the liquid, as compared to its optical refractive index and thermal conductance. On top of the above-mentioned problem of un-proper filling by the most viscous liquids of the volume below the disk, our earlier research \cite{neshasteh2023fluid} indicates that compressibility effects become significant at frequencies above 1 GHz, whereas within the operating frequency range of the device discussed here, compressibility accounts for only about $20\%$ of the mechanical effect of the liquid onto the disk. This aspect negatively impacts the accuracy of our approach when it comes to determine the compressibility (and speed of sound) of the liquid. Employing a disk with a taller pedestal and smaller radius, hence even higher mechanical frequency, could enhance the accuracy of this specific measurement. \\
Table 2 shows a comparison between the sensing results obtained with our approach and those obtained using other liquid sensing techniques. This comparison frames our findings within a broader context, illustrating the relative performances, advantages, and limitations of prior approaches. Our technique achieves accuracy levels comparable to prior techniques for each type of fluid measurement, but it is the only reported probe measuring all properties of interest of the fluid at the same time. This asset eliminates the need of multiple distinct technologies, and comes with an advantageous microscale footprint. 

%the developed optical, thermal, and mechanical model for a suspended disk (Eqs. \ref{eq:coupled1}-\ref{eq:coupled3}) using experimental data measured in water , 

% by comparing the results with available values from the literature (listed in Table \ref{tab:waterGlycerol}).

% \begin{table}
	%     \centering
	%     \begin{tabular}{cccccccc}
		%          &  $n$&  $\mu$ &  $k$& $c$  & $\rho$ &  & \\
		%          %  & refractive index & viscosity & thermal conductivity &  speed of sound& density & \\
		%           & & $(\text{m\,Pa\,s})$ & $(\text{W/m\,K})$& $(\text{km/s})$ & (g/cm$^3$) &\\% &  &  & \\
		%           this work &  &  &  &  &  & \\
		%            &  &  &  &  &  & \\
		%            &  &  &  &  &  & \\
		%     \end{tabular}
	%     \caption{Precision of measurement results using different technique}
	%     \label{tab:my_label}
	% \end{table}

% \begin{table}[h]
	% \centering
	% \caption{Add area and acquisition time, power ?? }
	% 	\label{tab:waterGlycerol}
	% 	\begin{tabular}{*8c}
		% 		\toprule
		% 		 & n & {$k \ (\frac{W}{m\cdot K})$}  &{$\rho \ (\frac{kg}{m^3})$}& {$\mu \ (mPa\cdot s)$} & {$c \ (\frac{m}{s})$} \\
		% 		\midrule
		% 		This work & 1.3180 & 0.5894& 997 & 1.05&1510\\
		% 		\cite{deng2013thermo} & 1.3474 & 0.5183& 1055&1.87&1728\\
		% 		  optical RI Plasmonic & 1.3759 & 0.4473& 1112&3.79&1844\\
		% 		   & 1.4050 & 0.3804& 1165&13.96&1909\\	
		% 		80 & 1.4319 & 0.3260& 1214&85.71&1919\\					
		% 		\bottomrule
		% 		\\
		% 		\\
		% \end{tabular}
	% \end{table}

As a last indicator of performance, we eventually investigate the minimum acquisition time of our disk sensor. The signal-to-noise ratio (SNR) of the RF detection unit utilized for measuring rheological properties is directly influenced by the optical power employed for the optomechanical measurement and by the integration time of the RF signal. In Fig. 4, the viscosity deduced from the analysis of mechanical frequency shift and quality factor is plotted against the optical power travelling in the bus waveguide and against the sweep time of the RF electronic spectrum analyser. The evolution of the spread of data in both panels of Fig. 4 indicate that a higher optical power and a longer integration time lead to better precision in viscosity determination.The viscosity determination can reach a precision better than $1\%$. Conversely, rapid measurements with a sweep time of $210\ \mu s$ result in a unity uncertainty in the determined value. This underscores the importance of trade-off between acquisition speed and measurement accuracy \cite{madsen2021ultrafast}. Our minimum measurement time is dictated by our electronic spectrum analyzer, and can be further reduced with a high-speed and low noise oscilloscope. Our current shortest measurement time (70 $\mu s$) comes close to the record value for viscosity measurement (20 $\mu s$) \cite{madsen2021ultrafast}, with an accuracy that is however twice better. Such acquisition times of few tens of microseconds are orders of magnitude shorter than those of conventional rheometry systems. For higher optical power, the error bars of our viscosity sensing platform diminish significantly. The detection error decreases from $10\%$ at $50\ \mu W$  power to below $4\%$ above $300\ \mu W$. Balancing the injected optical power level is crucial for accurate and reliable measurements of rheological properties of the liquid, while minimizing perturbation of the liquid temperature.  

%highlighting the critical role of injected power in improving measurement precision. This emphasizes the importance of optimizing injected power levels to achieve accurate and reliable measurements of rheological properties without affecting the temperature equilibrium.
\begin{figure}[h]%
	\centering
	\includegraphics[width=0.9\textwidth]{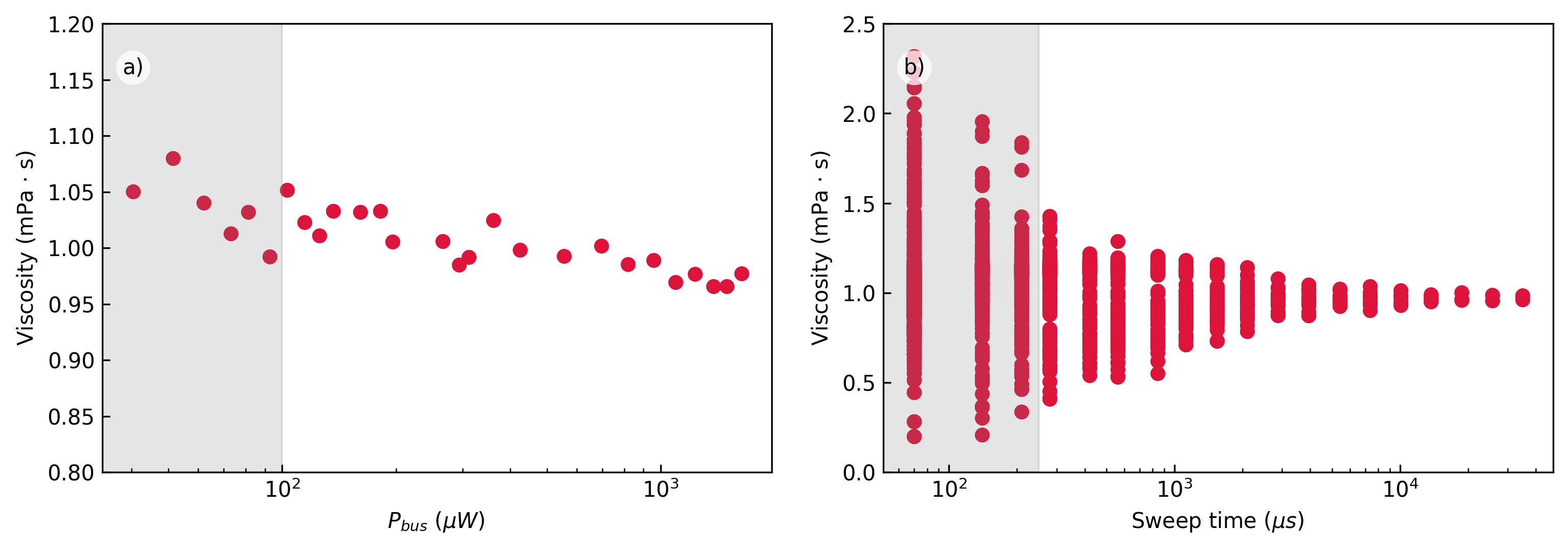}
	\caption{Extracted liquid viscosity, for varying operation conditions of the sensor platform. (a) Fixed sweep time (10 ms) and varying optical power travelling in the bus waveguide. (b) Fixed optical power in the bus waveguide ($800 \ \mu W$) and varying sweep time. Each red circle corresponds to one measured RF spectrum.}\label{fig4}
\end{figure}

% \begin{landscape}
	% \begin{figure*}[h]%
		% \centering
		% \includegraphics[width=1.6\textwidth]{table.pdf}
		% \caption{}\label{table1}
		% \end{figure*}
	% \end{landscape}

% \begin{table}[!htbp]
	% \centering
	% \caption{Comparison of percentages.}
	% \begin{tabular}{*5c}
		% \toprule
		% Mode &  \multicolumn{2}{c}{Var} & \multicolumn{2}{c}{Cum}\\
		% \midrule
		% {}   & EF   & CHF    & EF2   & CHF2\\
		% 1   &  17.5 & 19.1   & 17.5  & 19.1\\
		% 2   &  11.8 & 12.7   & 29.3  & 31.9\\
		% 3   &  6.6  &  5.6   & 35.9  & 37.4\\
		% \bottomrule
		% \end{tabular}
	% \end{table}
% Table generated by Excel2LaTeX from sheet 'Sheet3'
\begin{landscape}
	\begin{table}
		\centering
		\caption{State of the art fluid characterization sensors}
		\label{table2}
		\resizebox{1.5\textwidth}{!}{%
			\begin{tabular}{ccccccccccccccccc}
				%\multicolumn{11}{c}{Table I. Some of recent fluid characterization sensors}\\
				\toprule\multirow{2}{0.1cm}{\begin{sideways}Ref\end{sideways}} & \multicolumn{1}{c}{\multirow{2}{*}{\begin{sideways}Tech.\end{sideways}}} & \multirow{2}{*}{Size ($\mu m$)} & {\multirow{2}{*}{Mechanical frequency (kHz)}} & \multicolumn{2}{c}{Viscosity} & \multicolumn{2}{c}{Density} & \multicolumn{2}{c}{Speed of sound} & \multicolumn{2}{c}{Optical refractive index} & \multicolumn{2}{c}{Thermal conductance}\\
				& \multicolumn{1}{c}{} &       &       & \multicolumn{1}{c}{DR (mPa$\cdot$s)} & \multicolumn{1}{c}{Ac. (\%)} & DR(kg/m$^3$)    & Ac.(\%) & DR(m/s)    &  Ac. (\%) & DR (RIU)  &  Res. (RIU) & DR (W/m$\cdot$K)    &  Ac. (\%)\\
				\midrule
				\cite{payam2017simultaneous}     & \multicolumn{1}{c}{\multirow{6}[0]{0cm}{\tablefootnote{Laser readout of cantilever or membrane vibrations}}}  & \multicolumn{1}{c}{100} & \multicolumn{1}{c}{5, 40, 200} &       & 10  &       & 10 & \multicolumn{2}{c}{N/A} & \multicolumn{2}{c}{N/A}& \multicolumn{2}{c}{N/A}\\
				\cite{huang2022piezoelectric}    &      & \multicolumn{1}{c}{1382} & \multicolumn{1}{c}{100} & 0.3-2.5 & 8  & 650-1000 & 1  & \multicolumn{2}{c}{N/A} & \multicolumn{2}{c}{N/A}& \multicolumn{2}{c}{N/A}\\
				\cite{tiwari2023tip}    &  &        \multicolumn{1}{c}{1000} & \multicolumn{1}{c}{40} & 1-270 & 5  & \multicolumn{2}{c}{N/A} & \multicolumn{2}{c}{N/A}& \multicolumn{2}{c}{N/A}& \multicolumn{2}{c}{N/A} \\
				\cite{toledo20213d}    &   &     \multicolumn{1}{c}{16000} & \multicolumn{1}{c}{20}    & 1.8-2 & 5  & 980-1080 & 0.54  & \multicolumn{2}{c}{N/A} & \multicolumn{2}{c}{N/A}& \multicolumn{2}{c}{N/A}\\
				\cite{khan2013online}     &  &      \multicolumn{1}{c}{200} & \multicolumn{1}{c}{130-150} & 1-2   & 2.5 & 0-1000 & 0.1  & \multicolumn{2}{c}{N/A} & \multicolumn{2}{c}{N/A}& \multicolumn{2}{c}{N/A}\\
				\cite{oliva2019array}     & &      \multicolumn{1}{c}{1000} & \multicolumn{1}{c}{35} & 1-219 & 9.6  & 998-1235 & 1.5 & \multicolumn{2}{c}{N/A} &\multicolumn{2}{c}{N/A}&\multicolumn{2}{c}{N/A}\\
				\\
				\hline
				\\
				{\cite{ledesma2022single}} & \multicolumn{1}{c}{{\tablefootnote{Piezoelectric micromachined ultrasonic transducer}}} & \multicolumn{1}{c}{{80}} & \multicolumn{1}{c}{{1500-4000}} & \multicolumn{2}{c}{{only provides estimates}} & \multicolumn{2}{c}{3} & {1500-1900} & {5} & \multicolumn{2}{c}{N/A}& \multicolumn{2}{c}{N/A}\\ 
				\\ 
				\hline
				\\
				\cite{madsen2021ultrafast}     & \multicolumn{1}{c}{\tablefootnote{Tracking microparticles trapped in an optical tweezer}} &    {N/A}   &    {N/A}   & 0.5-2 & 4 ** & \multicolumn{2}{c}{N/A} & \multicolumn{2}{c}{N/A}& \multicolumn{2}{c}{N/A}& \multicolumn{2}{c}{N/A} \\
				\\
				\hline
				\\
				
				\cite{schmidt2008optical}     & \multicolumn{1}{c}{\tablefootnote{Optical pump-probe
						method}} &   {30}    & {N/A}      &\multicolumn{2}{c}{N/A} & \multicolumn{2}{c}{N/A} & \multicolumn{2}{c}{N/A}& \multicolumn{2}{c}{N/A}& {0.05-0.6}       & {5} \\
				\\
				\hline
				\\
				\cite{lazaro2021fbg}     & \multicolumn{1}{c}{\multirow{3}[0]{0cm}{\tablefootnote{Fiber Bragg
							Grating approach}}} &     10000  & {N/A}      &\multicolumn{2}{c}{N/A} &  {820-995} & 5 & \multicolumn{2}{c}{N/A}& \multicolumn{2}{c}{N/A}& {0.1-0.6}       & {7} \\
				\cite{ferreira2014refractive}     &  &     {N/A}  & 302.2, 400      &{1-2} & {3} &  \multicolumn{2}{c}{N/A} & \multicolumn{2}{c}{N/A}& {1.3-1.4} & {1$\times 10^{-4}$}& \multicolumn{2}{c}{N/A} \\
				\cite{pevec2017multiparameter}     &  &     3385 & {N/A}      &\multicolumn{2}{c}{N/A} &  \multicolumn{2}{c}{N/A} & \multicolumn{2}{c}{N/A}& {1.00012-1.00032} & {5$\times 10^{-7}$}& {0.03-0.18} & 3 \\
				\\
				\hline
				\\
				\multicolumn{1}{c}{{{\begin{sideways}This work\end{sideways}}}}& \multicolumn{1}{c}{{\tablefootnote{Suspended microdisk optomechanical resonator}}}& {16} & {300000} & {1-2} & {2} &{1000-1300} &{10} & {1500-1900}      & {10} &{1.3-1.5} & {4$\times 10^{-6}$} & {0.3-0.6}       & {2} \\ 
				\bottomrule
		\end{tabular}}%
	\end{table}
\end{landscape}

\section{Conclusion}\label{sec13}

Our study presents an optomechanical method for the measurement of the optical, thermal and rheological properties of liquids on the micron scale. By using a small suspended silicon disk as a passive probe, we achieve accurate measurements, while adding minimal disturbance to the investigated liquid. In particular, the use of a working power of $50 \ \mu W$, leading to a temperature change of less than $0.3^{\circ}$, preserves thermal equilibrium during measurements. By employing optomechanical transduction, we access to a wide range of fluid properties with a single device, including optical refractive index, thermal conductivity, viscosity, density and compressibility (hence speed of sound). With a reported fast acquisition time down to $70\ \mu s$, these properties can all be tracked with high time resolution. This opens applicability of the method to multiple scientific domains encompassing thermodynamical transitions in liquid, wetting phenomena, and in situ monitoring of biomolecule solutions.
\begin{acknowledgement}

\end{acknowledgement}
This work was supported by the European Research Council through CoG NOMLI (770933), by the Région Ile de France through the DIM-QuanTiP program, and by the Carnot network through the HERMES project.

\bibliography{achemso-demo}

\providecommand{\latin}[1]{#1}
\makeatletter
\providecommand{\doi}
  {\begingroup\let\do\@makeother\dospecials
  \catcode`\{=1 \catcode`\}=2 \doi@aux}
\providecommand{\doi@aux}[1]{\endgroup\texttt{#1}}
\makeatother
\providecommand*\mcitethebibliography{\thebibliography}
\csname @ifundefined\endcsname{endmcitethebibliography}
  {\let\endmcitethebibliography\endthebibliography}{}
\begin{mcitethebibliography}{49}
\providecommand*\natexlab[1]{#1}
\providecommand*\mciteSetBstSublistMode[1]{}
\providecommand*\mciteSetBstMaxWidthForm[2]{}
\providecommand*\mciteBstWouldAddEndPuncttrue
  {\def\EndOfBibitem{\unskip.}}
\providecommand*\mciteBstWouldAddEndPunctfalse
  {\let\EndOfBibitem\relax}
\providecommand*\mciteSetBstMidEndSepPunct[3]{}
\providecommand*\mciteSetBstSublistLabelBeginEnd[3]{}
\providecommand*\EndOfBibitem{}
\mciteSetBstSublistMode{f}
\mciteSetBstMaxWidthForm{subitem}{(\alph{mcitesubitemcount})}
\mciteSetBstSublistLabelBeginEnd
  {\mcitemaxwidthsubitemform\space}
  {\relax}
  {\relax}

\bibitem[Puneeth \latin{et~al.}(2021)Puneeth, Kulkarni, and
  Goel]{puneeth2021microfluidic}
Puneeth,~S.; Kulkarni,~M.~B.; Goel,~S. Microfluidic viscometers for biochemical
  and biomedical applications: A review. \emph{Engineering Research Express}
  \textbf{2021}, \emph{3}, 022003\relax
\mciteBstWouldAddEndPuncttrue
\mciteSetBstMidEndSepPunct{\mcitedefaultmidpunct}
{\mcitedefaultendpunct}{\mcitedefaultseppunct}\relax
\EndOfBibitem
\bibitem[Del~Giudice(2022)]{del2022review}
Del~Giudice,~F. A review of microfluidic devices for rheological
  characterisation. \emph{Micromachines} \textbf{2022}, \emph{13}, 167\relax
\mciteBstWouldAddEndPuncttrue
\mciteSetBstMidEndSepPunct{\mcitedefaultmidpunct}
{\mcitedefaultendpunct}{\mcitedefaultseppunct}\relax
\EndOfBibitem
\bibitem[Gu \latin{et~al.}(2023)Gu, Shanahan, Hart, Belser, Shofer, Atature,
  and Knowles]{gu2023simultaneous}
Gu,~Q.; Shanahan,~L.; Hart,~J.~W.; Belser,~S.; Shofer,~N.; Atature,~M.;
  Knowles,~H.~S. Simultaneous nanorheometry and nanothermometry using
  intracellular diamond quantum sensors. \emph{ACS nano} \textbf{2023},
  \emph{17}, 20034--20042\relax
\mciteBstWouldAddEndPuncttrue
\mciteSetBstMidEndSepPunct{\mcitedefaultmidpunct}
{\mcitedefaultendpunct}{\mcitedefaultseppunct}\relax
\EndOfBibitem
\bibitem[Salipante(2023)]{salipante2023microfluidic}
Salipante,~P.~F. Microfluidic techniques for mechanical measurements of
  biological samples. \emph{Biophysics Reviews} \textbf{2023}, \emph{4}\relax
\mciteBstWouldAddEndPuncttrue
\mciteSetBstMidEndSepPunct{\mcitedefaultmidpunct}
{\mcitedefaultendpunct}{\mcitedefaultseppunct}\relax
\EndOfBibitem
\bibitem[Singh \latin{et~al.}(2022)Singh, Sharma, Puchades, and
  Agarwal]{singh2022comprehensive}
Singh,~P.; Sharma,~K.; Puchades,~I.; Agarwal,~P.~B. A comprehensive review on
  MEMS-based viscometers. \emph{Sensors and Actuators A: Physical}
  \textbf{2022}, \emph{338}, 113456\relax
\mciteBstWouldAddEndPuncttrue
\mciteSetBstMidEndSepPunct{\mcitedefaultmidpunct}
{\mcitedefaultendpunct}{\mcitedefaultseppunct}\relax
\EndOfBibitem
\bibitem[Toropov \latin{et~al.}(2021)Toropov, Cabello, Serrano, Gutha, Rafti,
  and Vollmer]{toropov2021review}
Toropov,~N.; Cabello,~G.; Serrano,~M.~P.; Gutha,~R.~R.; Rafti,~M.; Vollmer,~F.
  Review of biosensing with whispering-gallery mode lasers. \emph{Light:
  Science \& Applications} \textbf{2021}, \emph{10}, 42\relax
\mciteBstWouldAddEndPuncttrue
\mciteSetBstMidEndSepPunct{\mcitedefaultmidpunct}
{\mcitedefaultendpunct}{\mcitedefaultseppunct}\relax
\EndOfBibitem
\bibitem[Souza \latin{et~al.}(2022)Souza, Faustino, Gon{\c{c}}alves, Moita,
  Ba{\~n}obre-L{\'o}pez, and Lima]{souza2022review}
Souza,~R.~R.; Faustino,~V.; Gon{\c{c}}alves,~I.~M.; Moita,~A.~S.;
  Ba{\~n}obre-L{\'o}pez,~M.; Lima,~R. A review of the advances and challenges
  in measuring the thermal conductivity of nanofluids. \emph{Nanomaterials}
  \textbf{2022}, \emph{12}, 2526\relax
\mciteBstWouldAddEndPuncttrue
\mciteSetBstMidEndSepPunct{\mcitedefaultmidpunct}
{\mcitedefaultendpunct}{\mcitedefaultseppunct}\relax
\EndOfBibitem
\bibitem[Zhou and Papautsky(2020)Zhou, and Papautsky]{zhou2020viscoelastic}
Zhou,~J.; Papautsky,~I. Viscoelastic microfluidics: Progress and challenges.
  \emph{Microsystems \& Nanoengineering} \textbf{2020}, \emph{6}, 113\relax
\mciteBstWouldAddEndPuncttrue
\mciteSetBstMidEndSepPunct{\mcitedefaultmidpunct}
{\mcitedefaultendpunct}{\mcitedefaultseppunct}\relax
\EndOfBibitem
\bibitem[Waigh(2016)]{waigh2016advances}
Waigh,~T.~A. Advances in the microrheology of complex fluids. \emph{Reports on
  Progress in Physics} \textbf{2016}, \emph{79}, 074601\relax
\mciteBstWouldAddEndPuncttrue
\mciteSetBstMidEndSepPunct{\mcitedefaultmidpunct}
{\mcitedefaultendpunct}{\mcitedefaultseppunct}\relax
\EndOfBibitem
\bibitem[Mao \latin{et~al.}(2022)Mao, Nielsen, and Ali]{mao2022passive}
Mao,~Y.; Nielsen,~P.; Ali,~J. Passive and active microrheology for biomedical
  systems. \emph{Frontiers in bioengineering and biotechnology} \textbf{2022},
  \emph{10}, 916354\relax
\mciteBstWouldAddEndPuncttrue
\mciteSetBstMidEndSepPunct{\mcitedefaultmidpunct}
{\mcitedefaultendpunct}{\mcitedefaultseppunct}\relax
\EndOfBibitem
\bibitem[Persson \latin{et~al.}(2020)Persson, Ambati, and
  Brandman]{persson2020cellular}
Persson,~L.~B.; Ambati,~V.~S.; Brandman,~O. Cellular control of viscosity
  counters changes in temperature and energy availability. \emph{Cell}
  \textbf{2020}, \emph{183}, 1572--1585\relax
\mciteBstWouldAddEndPuncttrue
\mciteSetBstMidEndSepPunct{\mcitedefaultmidpunct}
{\mcitedefaultendpunct}{\mcitedefaultseppunct}\relax
\EndOfBibitem
\bibitem[Budin \latin{et~al.}(2018)Budin, de~Rond, Chen, Chan, Petzold, and
  Keasling]{budin2018viscous}
Budin,~I.; de~Rond,~T.; Chen,~Y.; Chan,~L. J.~G.; Petzold,~C.~J.;
  Keasling,~J.~D. Viscous control of cellular respiration by membrane lipid
  composition. \emph{Science} \textbf{2018}, \emph{362}, 1186--1189\relax
\mciteBstWouldAddEndPuncttrue
\mciteSetBstMidEndSepPunct{\mcitedefaultmidpunct}
{\mcitedefaultendpunct}{\mcitedefaultseppunct}\relax
\EndOfBibitem
\bibitem[Trejo-Soto \latin{et~al.}(2022)Trejo-Soto, L{\'a}zaro, Pagonabarraga,
  and Hern{\'a}ndez-Machado]{trejo2022microfluidics}
Trejo-Soto,~C.; L{\'a}zaro,~G.~R.; Pagonabarraga,~I.; Hern{\'a}ndez-Machado,~A.
  Microfluidics approach to the mechanical properties of red blood cell
  membrane and their effect on blood rheology. \emph{Membranes} \textbf{2022},
  \emph{12}, 217\relax
\mciteBstWouldAddEndPuncttrue
\mciteSetBstMidEndSepPunct{\mcitedefaultmidpunct}
{\mcitedefaultendpunct}{\mcitedefaultseppunct}\relax
\EndOfBibitem
\bibitem[Robin \latin{et~al.}(2023)Robin, Emmerich, Ismail, Nigu{\`e}s, You,
  Nam, Keerthi, Siria, Geim, Radha, \latin{et~al.} others]{robin2023long}
Robin,~P.; Emmerich,~T.; Ismail,~A.; Nigu{\`e}s,~A.; You,~Y.; Nam,~G.-H.;
  Keerthi,~A.; Siria,~A.; Geim,~A.; Radha,~B.; others Long-term memory and
  synapse-like dynamics in two-dimensional nanofluidic channels. \emph{Science}
  \textbf{2023}, \emph{379}, 161--167\relax
\mciteBstWouldAddEndPuncttrue
\mciteSetBstMidEndSepPunct{\mcitedefaultmidpunct}
{\mcitedefaultendpunct}{\mcitedefaultseppunct}\relax
\EndOfBibitem
\bibitem[Xiong \latin{et~al.}(2023)Xiong, Li, He, Xie, Zong, Jiang, Ma, Wu,
  Fei, Yu, \latin{et~al.} others]{xiong2023neuromorphic}
Xiong,~T.; Li,~C.; He,~X.; Xie,~B.; Zong,~J.; Jiang,~Y.; Ma,~W.; Wu,~F.;
  Fei,~J.; Yu,~P.; others Neuromorphic functions with a
  polyelectrolyte-confined fluidic memristor. \emph{Science} \textbf{2023},
  \emph{379}, 156--161\relax
\mciteBstWouldAddEndPuncttrue
\mciteSetBstMidEndSepPunct{\mcitedefaultmidpunct}
{\mcitedefaultendpunct}{\mcitedefaultseppunct}\relax
\EndOfBibitem
\bibitem[Hou \latin{et~al.}(2017)Hou, Zhang, Santiago, Alvarez, Ribas, Jonas,
  Weiss, Andrews, Aizenberg, and Khademhosseini]{hou2017interplay}
Hou,~X.; Zhang,~Y.~S.; Santiago,~G. T.-d.; Alvarez,~M.~M.; Ribas,~J.;
  Jonas,~S.~J.; Weiss,~P.~S.; Andrews,~A.~M.; Aizenberg,~J.; Khademhosseini,~A.
  Interplay between materials and microfluidics. \emph{Nature Reviews
  Materials} \textbf{2017}, \emph{2}, 1--15\relax
\mciteBstWouldAddEndPuncttrue
\mciteSetBstMidEndSepPunct{\mcitedefaultmidpunct}
{\mcitedefaultendpunct}{\mcitedefaultseppunct}\relax
\EndOfBibitem
\bibitem[Liu and Jiang(2017)Liu, and Jiang]{liu2017microfluidics}
Liu,~Y.; Jiang,~X. Why microfluidics? Merits and trends in chemical synthesis.
  \emph{Lab on a Chip} \textbf{2017}, \emph{17}, 3960--3978\relax
\mciteBstWouldAddEndPuncttrue
\mciteSetBstMidEndSepPunct{\mcitedefaultmidpunct}
{\mcitedefaultendpunct}{\mcitedefaultseppunct}\relax
\EndOfBibitem
\bibitem[Choi and Park(2010)Choi, and Park]{choi2010microfluidic}
Choi,~S.; Park,~J.-K. Microfluidic rheometer for characterization of protein
  unfolding and aggregation in microflows. \emph{Small} \textbf{2010},
  \emph{6}, 1306--1310\relax
\mciteBstWouldAddEndPuncttrue
\mciteSetBstMidEndSepPunct{\mcitedefaultmidpunct}
{\mcitedefaultendpunct}{\mcitedefaultseppunct}\relax
\EndOfBibitem
\bibitem[Sbarra \latin{et~al.}(2022)Sbarra, Waquier, Suffit, Lema{\^\i}tre, and
  Favero]{Samantha2022}
Sbarra,~S.; Waquier,~L.; Suffit,~S.; Lema{\^\i}tre,~A.; Favero,~I.
  Optomechanical measurement of single nanodroplet evaporation with millisecond
  time-resolution. \emph{Nature Communications} \textbf{2022}, \emph{13},
  6462\relax
\mciteBstWouldAddEndPuncttrue
\mciteSetBstMidEndSepPunct{\mcitedefaultmidpunct}
{\mcitedefaultendpunct}{\mcitedefaultseppunct}\relax
\EndOfBibitem
\bibitem[Jimenez \latin{et~al.}(2023)Jimenez, Creton, Marliere, Teule-Gay,
  Nguyen, and Marre]{jimenez2023microfluidic}
Jimenez,~R.~M.; Creton,~B.; Marliere,~C.; Teule-Gay,~L.; Nguyen,~O.; Marre,~S.
  A microfluidic strategy for accessing the thermal conductivity of liquids at
  different temperatures. \emph{Microchemical Journal} \textbf{2023},
  \emph{193}, 109030\relax
\mciteBstWouldAddEndPuncttrue
\mciteSetBstMidEndSepPunct{\mcitedefaultmidpunct}
{\mcitedefaultendpunct}{\mcitedefaultseppunct}\relax
\EndOfBibitem
\bibitem[Payam \latin{et~al.}(2017)Payam, Trewby, and
  Vo{\"\i}tchovsky]{payam2017simultaneous}
Payam,~A.~F.; Trewby,~W.; Vo{\"\i}tchovsky,~K. Simultaneous viscosity and
  density measurement of small volumes of liquids using a vibrating
  microcantilever. \emph{Analyst} \textbf{2017}, \emph{142}, 1492--1498\relax
\mciteBstWouldAddEndPuncttrue
\mciteSetBstMidEndSepPunct{\mcitedefaultmidpunct}
{\mcitedefaultendpunct}{\mcitedefaultseppunct}\relax
\EndOfBibitem
\bibitem[Huang \latin{et~al.}(2022)Huang, Li, Luo, Lu, Zhao, Yang, Wang, Wang,
  Lin, and Jiang]{huang2022piezoelectric}
Huang,~L.; Li,~W.; Luo,~G.; Lu,~D.; Zhao,~L.; Yang,~P.; Wang,~X.; Wang,~J.;
  Lin,~Q.; Jiang,~Z. Piezoelectric-AlN resonators at two-dimensional flexural
  modes for the density and viscosity decoupled determination of liquids.
  \emph{Microsystems \& nanoengineering} \textbf{2022}, \emph{8}, 38\relax
\mciteBstWouldAddEndPuncttrue
\mciteSetBstMidEndSepPunct{\mcitedefaultmidpunct}
{\mcitedefaultendpunct}{\mcitedefaultseppunct}\relax
\EndOfBibitem
\bibitem[Tiwari \latin{et~al.}(2023)Tiwari, Dangi, and Pratap]{tiwari2023tip}
Tiwari,~S.; Dangi,~A.; Pratap,~R. A tip-coupled, two-cantilever, non-resonant
  microsystem for direct measurement of liquid viscosity. \emph{Microsystems \&
  nanoengineering} \textbf{2023}, \emph{9}, 34\relax
\mciteBstWouldAddEndPuncttrue
\mciteSetBstMidEndSepPunct{\mcitedefaultmidpunct}
{\mcitedefaultendpunct}{\mcitedefaultseppunct}\relax
\EndOfBibitem
\bibitem[Toledo \latin{et~al.}(2021)Toledo, Ruiz-D{\'\i}ez, Velasco,
  Hernando-Garc{\'\i}a, and S{\'a}nchez-Rojas]{toledo20213d}
Toledo,~J.; Ruiz-D{\'\i}ez,~V.; Velasco,~J.; Hernando-Garc{\'\i}a,~J.;
  S{\'a}nchez-Rojas,~J.~L. 3D-printed liquid cell resonator with piezoelectric
  actuation for in-line density-viscosity measurements. \emph{Sensors}
  \textbf{2021}, \emph{21}, 7654\relax
\mciteBstWouldAddEndPuncttrue
\mciteSetBstMidEndSepPunct{\mcitedefaultmidpunct}
{\mcitedefaultendpunct}{\mcitedefaultseppunct}\relax
\EndOfBibitem
\bibitem[Khan \latin{et~al.}(2013)Khan, Schmid, Larsen, Davis, Yan, Stenby, and
  Boisen]{khan2013online}
Khan,~M.; Schmid,~S.; Larsen,~P.~E.; Davis,~Z.~J.; Yan,~W.; Stenby,~E.~H.;
  Boisen,~A. Online measurement of mass density and viscosity of pL fluid
  samples with suspended microchannel resonator. \emph{Sensors and Actuators B:
  Chemical} \textbf{2013}, \emph{185}, 456--461\relax
\mciteBstWouldAddEndPuncttrue
\mciteSetBstMidEndSepPunct{\mcitedefaultmidpunct}
{\mcitedefaultendpunct}{\mcitedefaultseppunct}\relax
\EndOfBibitem
\bibitem[Oliva \latin{et~al.}(2019)Oliva, Bircher, Schoenenberger, and
  Braun]{oliva2019array}
Oliva,~P.; Bircher,~B.~A.; Schoenenberger,~C.-A.; Braun,~T. Array based
  real-time measurement of fluid viscosities and mass-densities to monitor
  biological filament formation. \emph{Lab on a Chip} \textbf{2019}, \emph{19},
  1305--1314\relax
\mciteBstWouldAddEndPuncttrue
\mciteSetBstMidEndSepPunct{\mcitedefaultmidpunct}
{\mcitedefaultendpunct}{\mcitedefaultseppunct}\relax
\EndOfBibitem
\bibitem[Ledesma \latin{et~al.}(2022)Ledesma, Zamora, Yanez, Uranga, and
  Barniol]{ledesma2022single}
Ledesma,~E.; Zamora,~I.; Yanez,~J.; Uranga,~A.; Barniol,~N. Single-cell system
  using monolithic PMUTs-on-CMOS to monitor fluid hydrodynamic properties.
  \emph{Microsystems \& Nanoengineering} \textbf{2022}, \emph{8}, 76\relax
\mciteBstWouldAddEndPuncttrue
\mciteSetBstMidEndSepPunct{\mcitedefaultmidpunct}
{\mcitedefaultendpunct}{\mcitedefaultseppunct}\relax
\EndOfBibitem
\bibitem[Madsen \latin{et~al.}(2021)Madsen, Waleed, Casacio, Terrasson,
  Stilgoe, Taylor, and Bowen]{madsen2021ultrafast}
Madsen,~L.~S.; Waleed,~M.; Casacio,~C.~A.; Terrasson,~A.; Stilgoe,~A.~B.;
  Taylor,~M.~A.; Bowen,~W.~P. Ultrafast viscosity measurement with ballistic
  optical tweezers. \emph{Nature photonics} \textbf{2021}, \emph{15},
  386--392\relax
\mciteBstWouldAddEndPuncttrue
\mciteSetBstMidEndSepPunct{\mcitedefaultmidpunct}
{\mcitedefaultendpunct}{\mcitedefaultseppunct}\relax
\EndOfBibitem
\bibitem[Neshasteh \latin{et~al.}(2024)Neshasteh, Shlesinger, Ravaro, G{\'e}ly,
  Jourdan, Hentz, and Favero]{neshasteh2024optomechanical}
Neshasteh,~H.; Shlesinger,~I.; Ravaro,~M.; G{\'e}ly,~M.; Jourdan,~G.;
  Hentz,~S.; Favero,~I. Optomechanical micro-rheology of complex fluids at
  ultra-high frequency. \emph{arXiv preprint arXiv:2410.13467} \textbf{2024},
  \relax
\mciteBstWouldAddEndPunctfalse
\mciteSetBstMidEndSepPunct{\mcitedefaultmidpunct}
{}{\mcitedefaultseppunct}\relax
\EndOfBibitem
\bibitem[Gil-Santos \latin{et~al.}(2015)Gil-Santos, Baker, Nguyen, Hease,
  Gomez, Lema{\^\i}tre, Ducci, Leo, and Favero]{gil2015high}
Gil-Santos,~E.; Baker,~C.; Nguyen,~D.; Hease,~W.; Gomez,~C.; Lema{\^\i}tre,~A.;
  Ducci,~S.; Leo,~G.; Favero,~I. High-frequency nano-optomechanical disk
  resonators in liquids. \emph{Nature nanotechnology} \textbf{2015}, \emph{10},
  810--816\relax
\mciteBstWouldAddEndPuncttrue
\mciteSetBstMidEndSepPunct{\mcitedefaultmidpunct}
{\mcitedefaultendpunct}{\mcitedefaultseppunct}\relax
\EndOfBibitem
\bibitem[Neshasteh \latin{et~al.}(2023)Neshasteh, Ravaro, and
  Favero]{neshasteh2023fluid}
Neshasteh,~H.; Ravaro,~M.; Favero,~I. Fluid--structure model for disks
  vibrating at ultra-high frequency in a compressible viscous fluid.
  \emph{Physics of Fluids} \textbf{2023}, \emph{35}\relax
\mciteBstWouldAddEndPuncttrue
\mciteSetBstMidEndSepPunct{\mcitedefaultmidpunct}
{\mcitedefaultendpunct}{\mcitedefaultseppunct}\relax
\EndOfBibitem
\bibitem[Sbarra \latin{et~al.}(2021)Sbarra, Allain, Suffit, Lema{\^\i}tre, and
  Favero]{Samantha2021}
Sbarra,~S.; Allain,~P.~E.; Suffit,~S.; Lema{\^\i}tre,~A.; Favero,~I. A
  multiphysics model for ultra-high frequency optomechanical resonators
  optically actuated and detected in the oscillating mode. \emph{APL Photonics}
  \textbf{2021}, \emph{6}, 8\relax
\mciteBstWouldAddEndPuncttrue
\mciteSetBstMidEndSepPunct{\mcitedefaultmidpunct}
{\mcitedefaultendpunct}{\mcitedefaultseppunct}\relax
\EndOfBibitem
\bibitem[Hale and Querry(1973)Hale, and Querry]{hale1973optical}
Hale,~G.~M.; Querry,~M.~R. Optical constants of water in the 200-nm to
  200-$\mu$m wavelength region. \emph{Applied optics} \textbf{1973}, \emph{12},
  555--563\relax
\mciteBstWouldAddEndPuncttrue
\mciteSetBstMidEndSepPunct{\mcitedefaultmidpunct}
{\mcitedefaultendpunct}{\mcitedefaultseppunct}\relax
\EndOfBibitem
\bibitem[gly(1963)]{glycerine1963physical}
\emph{Physical properties of glycerine and its solutions}; Glycerine Producers'
  Association, 1963\relax
\mciteBstWouldAddEndPuncttrue
\mciteSetBstMidEndSepPunct{\mcitedefaultmidpunct}
{\mcitedefaultendpunct}{\mcitedefaultseppunct}\relax
\EndOfBibitem
\bibitem[Bates(1936)]{bates1936binary}
Bates,~O.~K. Binary mixtures of water and glycerol-thermal conductivity of
  liquids. \emph{Industrial \& Engineering Chemistry} \textbf{1936}, \emph{28},
  494--498\relax
\mciteBstWouldAddEndPuncttrue
\mciteSetBstMidEndSepPunct{\mcitedefaultmidpunct}
{\mcitedefaultendpunct}{\mcitedefaultseppunct}\relax
\EndOfBibitem
\bibitem[Volk and K{\"a}hler(2018)Volk, and K{\"a}hler]{volk2018density}
Volk,~A.; K{\"a}hler,~C.~J. Density model for aqueous glycerol solutions.
  \emph{Experiments in Fluids} \textbf{2018}, \emph{59}, 75\relax
\mciteBstWouldAddEndPuncttrue
\mciteSetBstMidEndSepPunct{\mcitedefaultmidpunct}
{\mcitedefaultendpunct}{\mcitedefaultseppunct}\relax
\EndOfBibitem
\bibitem[Cheng(2008)]{cheng2008formula}
Cheng,~N.-S. Formula for the viscosity of a glycerol- water mixture.
  \emph{Industrial \& engineering chemistry research} \textbf{2008}, \emph{47},
  3285--3288\relax
\mciteBstWouldAddEndPuncttrue
\mciteSetBstMidEndSepPunct{\mcitedefaultmidpunct}
{\mcitedefaultendpunct}{\mcitedefaultseppunct}\relax
\EndOfBibitem
\bibitem[Slie \latin{et~al.}(1966)Slie, Donfor~Jr, and
  Litovitz]{slie1966ultrasonic}
Slie,~W.; Donfor~Jr,~A.; Litovitz,~T. Ultrasonic shear and longitudinal
  measurements in aqueous glycerol. \emph{The Journal of Chemical Physics}
  \textbf{1966}, \emph{44}, 3712--3718\relax
\mciteBstWouldAddEndPuncttrue
\mciteSetBstMidEndSepPunct{\mcitedefaultmidpunct}
{\mcitedefaultendpunct}{\mcitedefaultseppunct}\relax
\EndOfBibitem
\bibitem[Parrain \latin{et~al.}(2015)Parrain, Baker, Wang, Guha, Santos,
  Lemaitre, Senellart, Leo, Ducci, and Favero]{parrain2015origin}
Parrain,~D.; Baker,~C.; Wang,~G.; Guha,~B.; Santos,~E.~G.; Lemaitre,~A.;
  Senellart,~P.; Leo,~G.; Ducci,~S.; Favero,~I. Origin of optical losses in
  gallium arsenide disk whispering gallery resonators. \emph{Optics express}
  \textbf{2015}, \emph{23}, 19656--19672\relax
\mciteBstWouldAddEndPuncttrue
\mciteSetBstMidEndSepPunct{\mcitedefaultmidpunct}
{\mcitedefaultendpunct}{\mcitedefaultseppunct}\relax
\EndOfBibitem
\bibitem[Neshasteh \latin{et~al.}(2015)Neshasteh, Mataji-Kojouri,
  Akbarzadeh-Jahromi, and Shahabadi]{neshasteh2015hybrid}
Neshasteh,~H.; Mataji-Kojouri,~A.; Akbarzadeh-Jahromi,~S.-A.; Shahabadi,~M. A
  hybrid photonic-plasmonic sensing platform for differentiating background and
  surface interactions using an array of metal-insulator-metal resonators.
  \emph{IEEE Sensors Journal} \textbf{2015}, \emph{16}, 1621--1627\relax
\mciteBstWouldAddEndPuncttrue
\mciteSetBstMidEndSepPunct{\mcitedefaultmidpunct}
{\mcitedefaultendpunct}{\mcitedefaultseppunct}\relax
\EndOfBibitem
\bibitem[Mataji-Kojouri \latin{et~al.}(2020)Mataji-Kojouri, Ozen, Shahabadi,
  Inci, and Demirci]{mataji2020entangled}
Mataji-Kojouri,~A.; Ozen,~M.~O.; Shahabadi,~M.; Inci,~F.; Demirci,~U. Entangled
  nanoplasmonic cavities for estimating thickness of surface-adsorbed layers.
  \emph{ACS nano} \textbf{2020}, \emph{14}, 8518--8527\relax
\mciteBstWouldAddEndPuncttrue
\mciteSetBstMidEndSepPunct{\mcitedefaultmidpunct}
{\mcitedefaultendpunct}{\mcitedefaultseppunct}\relax
\EndOfBibitem
\bibitem[Vollmer and Yang(2012)Vollmer, and Yang]{vollmer2012review}
Vollmer,~F.; Yang,~L. Review Label-free detection with high-Q microcavities: a
  review of biosensing mechanisms for integrated devices. \emph{Nanophotonics}
  \textbf{2012}, \emph{1}, 267--291\relax
\mciteBstWouldAddEndPuncttrue
\mciteSetBstMidEndSepPunct{\mcitedefaultmidpunct}
{\mcitedefaultendpunct}{\mcitedefaultseppunct}\relax
\EndOfBibitem
\bibitem[Guha \latin{et~al.}(2017)Guha, Mariani, Lema{\^\i}tre, Combri{\'e},
  Leo, and Favero]{guha2017high}
Guha,~B.; Mariani,~S.; Lema{\^\i}tre,~A.; Combri{\'e},~S.; Leo,~G.; Favero,~I.
  High frequency optomechanical disk resonators in III--V ternary
  semiconductors. \emph{Optics express} \textbf{2017}, \emph{25},
  24639--24649\relax
\mciteBstWouldAddEndPuncttrue
\mciteSetBstMidEndSepPunct{\mcitedefaultmidpunct}
{\mcitedefaultendpunct}{\mcitedefaultseppunct}\relax
\EndOfBibitem
\bibitem[You \latin{et~al.}(2013)You, Yun, and Lee]{you2013surface}
You,~I.; Yun,~N.; Lee,~H. Surface-Tension-Confined Microfluidics and Their
  Applications. \emph{ChemPhysChem} \textbf{2013}, \emph{14}, 471--481\relax
\mciteBstWouldAddEndPuncttrue
\mciteSetBstMidEndSepPunct{\mcitedefaultmidpunct}
{\mcitedefaultendpunct}{\mcitedefaultseppunct}\relax
\EndOfBibitem
\bibitem[Schmidt \latin{et~al.}(2008)Schmidt, Chiesa, Chen, and
  Chen]{schmidt2008optical}
Schmidt,~A.; Chiesa,~M.; Chen,~X.; Chen,~G. An optical pump-probe technique for
  measuring the thermal conductivity of liquids. \emph{Review of Scientific
  Instruments} \textbf{2008}, \emph{79}\relax
\mciteBstWouldAddEndPuncttrue
\mciteSetBstMidEndSepPunct{\mcitedefaultmidpunct}
{\mcitedefaultendpunct}{\mcitedefaultseppunct}\relax
\EndOfBibitem
\bibitem[Lazaro \latin{et~al.}(2021)Lazaro, Marques, Castellani, and
  Leal-Junior]{lazaro2021fbg}
Lazaro,~R.~C.; Marques,~C.; Castellani,~C.~E.; Leal-Junior,~A. Fbg-based
  measurement systems for density, specific heat capacity and thermal
  conductivity assessment for liquids. \emph{IEEE Sensors Journal}
  \textbf{2021}, \emph{21}, 7657--7664\relax
\mciteBstWouldAddEndPuncttrue
\mciteSetBstMidEndSepPunct{\mcitedefaultmidpunct}
{\mcitedefaultendpunct}{\mcitedefaultseppunct}\relax
\EndOfBibitem
\bibitem[Ferreira \latin{et~al.}(2014)Ferreira, Bilro, Marques, Oliveira, and
  Nogueira]{ferreira2014refractive}
Ferreira,~R.; Bilro,~L.; Marques,~C.; Oliveira,~R.; Nogueira,~R. Refractive
  index and viscosity: dual sensing with plastic fibre gratings. 23rd
  International Conference on Optical Fibre Sensors. 2014; pp 1297--1300\relax
\mciteBstWouldAddEndPuncttrue
\mciteSetBstMidEndSepPunct{\mcitedefaultmidpunct}
{\mcitedefaultendpunct}{\mcitedefaultseppunct}\relax
\EndOfBibitem
\bibitem[Pevec and Donlagic(2017)Pevec, and Donlagic]{pevec2017multiparameter}
Pevec,~S.; Donlagic,~D. Multiparameter fiber-optic sensor for simultaneous
  measurement of thermal conductivity, pressure, refractive index, and
  temperature. \emph{IEEE Photonics Journal} \textbf{2017}, \emph{9},
  1--14\relax
\mciteBstWouldAddEndPuncttrue
\mciteSetBstMidEndSepPunct{\mcitedefaultmidpunct}
{\mcitedefaultendpunct}{\mcitedefaultseppunct}\relax
\EndOfBibitem
\end{mcitethebibliography}

\end{document}